\documentclass[sn-nature]{sn-jnl}
\usepackage{graphicx}%
\usepackage{multirow}%
\usepackage{amsmath,amssymb,amsfonts}%
\usepackage{amsthm}%
\usepackage{mathrsfs}%
\usepackage[title]{appendix}%
\usepackage{xcolor}%
\usepackage{textcomp}%
\usepackage{manyfoot}%
\usepackage{booktabs}%
\usepackage{algpseudocode}%
\usepackage{listings}%
\usepackage{geometry}
\usepackage{ulem}
 
\def\be{\begin{equation}}
\def\ee{\end{equation}}
\def\bea{\begin{eqnarray}}
\def\eea{\end{eqnarray}}

\begin{document}
\title{\textbf{{ Quantum Force Sensing by Digital Twinning of Atomic Bose-Einstein Condensates}}} 
	
	\author[1,2]{\fnm{Tangyou} \sur{Huang}}
	
	\author[3]{\fnm{Zhongcheng} \sur{Yu}}
	
	\author[1,2]{\fnm{Zhongyi} \sur{Ni}}
	
	\author[3,4,5]{\fnm{Xiaoji} \sur{Zhou}}
	
	\author*[2,6,1,7,8]{\fnm{Xiaopeng}       \sur{Li}}\email{xiaopeng\_li@fudan.edu.cn}
	
	\affil[1]{\orgdiv{Shanghai Qi Zhi Institute}, \orgname{AI Tower}, \orgaddress{\street{Xuhui District}, \city{Shanghai}, \postcode{200232},  \country{China}}}
	
	\affil[2]{State Key Laboratory of Surface Physics, Department of Physics, Fudan University, Shanghai 200433, China}
	\affil[3]{State Key Laboratory of Advanced Optical Communication System and Network, School of Electronics, Peking University, Beijing 100871, China}
	
	\affil[4]{Institute of Advanced Functional Materials and Devices, Shanxi University, Taiyuan 030031, China}
	\affil[5]{Institute of carbon-based thin film electronics, Peking University, Shanxi, Taiyuan  030012,China }

	\affil[6]{Institute for Nanoelectronic Devices and Quantum Computing, Fudan University, Shanghai 200433, China}
	\affil[7]{Shanghai Artificial Intelligience Laboratory, Shanghai 200232, China} 
	\affil[8]{Shanghai Research Center for Quantum Sciences, Shanghai 201315, China}

	\abstract{
High sensitivity detection plays a vital role in  science discoveries and technological applications.  
While intriguing methods utilizing collective many-body correlations and quantum entanglements have been developed in physics to enhance sensitivity, their practical implementation remains challenging due to rigorous technological requirements. 
Here, we propose an entirely data-driven approach that harnesses the capabilities of machine learning, to significantly augment weak-signal detection sensitivity. 
{ In an atomic force sensor, our method combines a digital replica of force-free data with anomaly detection technique, devoid of any prior knowledge about the physical system or assumptions regarding the sensing process. }
Our findings demonstrate a significant advancement in sensitivity, achieving an order of magnitude improvement over conventional protocols in detecting a weak force of approximately $10^{-25}~\mathrm{N}$. The resulting sensitivity reaches 
$1.7(4) \times 10^{-25}~\mathrm{N}/\sqrt{\mathrm{Hz}}$. Our machine learning-based signal processing approach does not rely on system-specific details or processed signals, rendering it highly applicable to sensing technologies across various domains.
		}
	\maketitle
	\newpage

	\section*{Introduction}\label{sec1}
 
	In recent decades, quantum technologies have made remarkable strides, culminating in the emergence of quantum sensing techniques that enable high-precision detection at the microscopic level. Quantum sensor leverages quantum resources to detect subtle changes in physical quantities such as time, force, and electromagnetic fields, providing extreme precision at the atomic scale \cite{atomicClock,schreppler2014optically,PhysRevX.12.021061}.
	These implementations have been successfully realized across diverse platforms, including cold atoms \cite{facon2016sensitive,Szigeti2020prl}, superconducting circuits \cite{halbertal2016nanoscale}, and solid-state spin systems \cite{taylor2008high}. Recently, progresses in quantum sensing highlight its capability for real-world applications such as precision navigation \cite{bidel2018absolute}, gravity detection \cite{stray2022gravity}, dark matter searches \cite{darkmatter} and others \cite{bongs2019taking,corre2017prx}.
  { However,  they are often facing challenges associated with both hardware setup and software processing, limiting the practical applications. 
 While hardware upgrades have demonstrated the capability to enhance
 sensing performance~\cite{mason2019continuous, corre2017prx, LIGO2020quantum,BeyondSQL}, software approaches like information compilation and data analysis offer an alternative path to improve the sensitivity of high-precision detection.}
 {In this work, we are dedicated to improving sensitivity through a machine-learning-assisted method, emphasizing the crucial role of advanced information processing in sensing applications.}

	Earlier computational approaches have {employed} statistical learning of signal acquisition from sensing observable taking into account the inherent noise and variations \cite{pertsinidis2010subnanometre,ligo2011gravitational,blums2018single}. In recent years,  machine learning approaches have been used for signal selection \cite{Colgan2020prd, stray2022gravity} and {sensing-related analysis with anomaly detection \cite{AD2018prl,AD2022prd}}.
	Nevertheless, these methods often demand a substantial amount of sensing data or rely on prior knowledge of signal and noise properties, thereby {still} limiting their applicability.
	More recently, the concept of digital twinning has gained significant attention as a powerful tool for simulating and understanding complex physical systems. Digital twinning involves creating a virtual replica or model that mirrors the behavior and properties of a physical system, allowing for real-time analysis, optimization, and prediction. 
	By effectively bridging the physical and virtual realms, digital twinning provides a unique opportunity to {improve the overall performance of quantum sensing application.}

	In this article, we propose an application of digital twinning for quantum force sensing in atomic Bose-Einstein condensates (BECs)~\cite{GUO2022Bulletin}. Leveraging the advancements in generative machine learning models, we create a digital twin that faithfully represents the atomic BEC system under investigation. 
	{The digital twin efficiently captures time-of-flight features affected by intricate correlations and non-linear dynamics of the physical system, enabling us to devise a data-driven approach for quantum force sensing based on anomaly detection.} 
	Conventional quantum force sensing techniques often rely on extracting basic statistical moments from high-dimensional experimental data, {with the information hidden in the high moments wasted}.
	In contrast, our digital twinning approach incorporates a generative machine learning model to construct a nonlinear function,
	 which takes advantage of the {valuable information hidden} 
	in the high-dimensional data. 
	This approach allows for improved signal-to-noise ratio and enhanced sensitivity, without compromising the long-term stability of the sensing system. We anticipate that our anomaly detection technique, facilitated by digital twinning, can be broadly applied to sensing experiments involving high-dimensional data acquisition cycles. By maximizing the utilization of high-dimensional data, our approach surpasses conventional techniques that rely solely on basic statistical features, unlocking new avenues for quantum force sensing and precision measurements.

	\medskip
	\section*{Results}\label{sec2}
	  
\begin{figure}[h]%
\centering
\includegraphics[ width=1\linewidth]{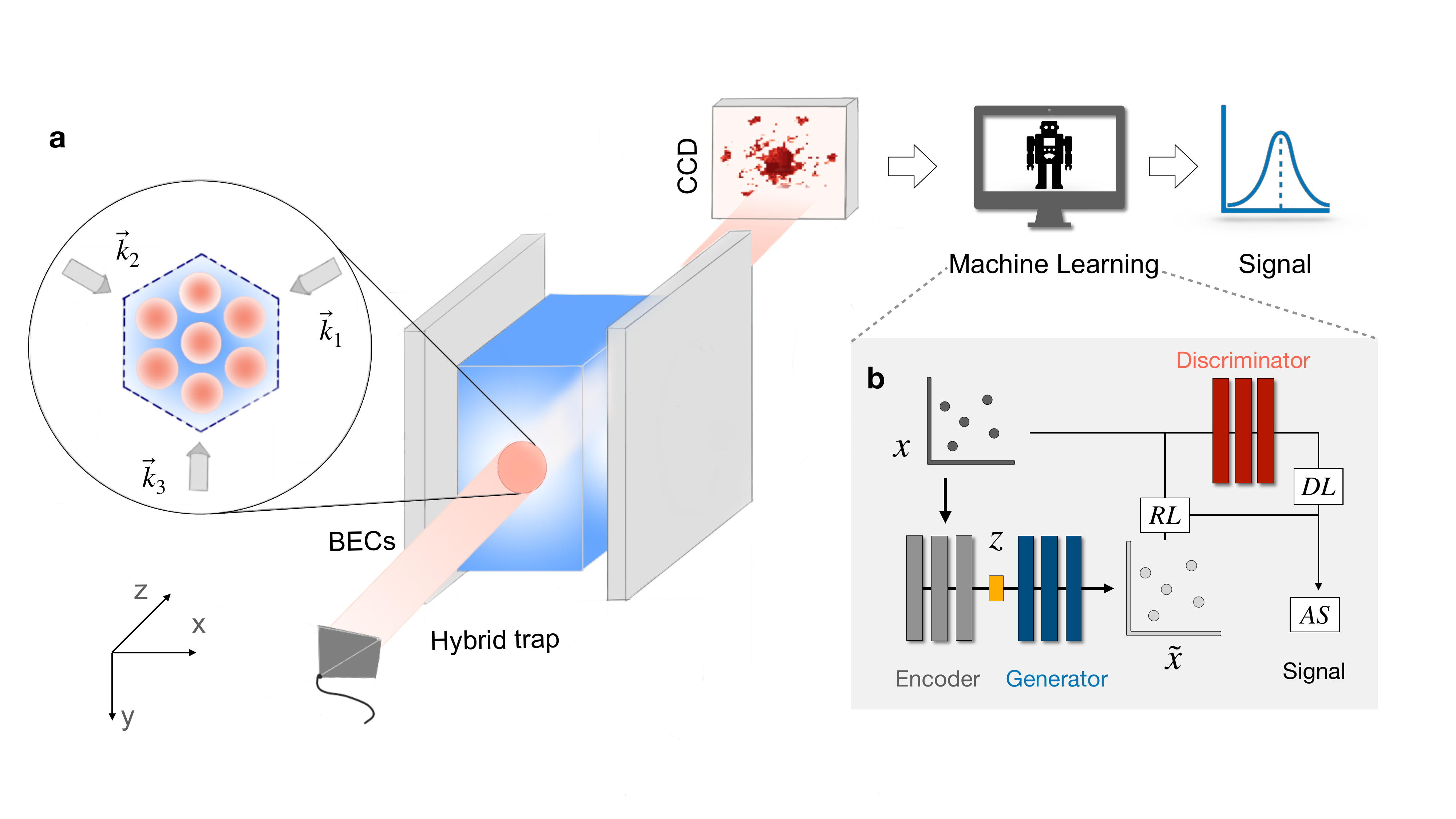}
\caption{
	\textbf{The machine-learning assisted atomic force sensing}. \textbf{a}. The schematic diagram of experiment setups for an atomic Bose-Einstein condensates (BEC) based force sensor. The atomic BEC is confined by  a triangular optical lattice spread in the x-y plane. The time-of-flight image is probed by an imaging laser beam along the $z$ direction, and showing  on the  CCD camera. With an external force applied on the BEC, the time-of-flight (TOF) image would develop  systematic shifts. The signal can be automatically generated by inputting the TOF image into the well-trained machine-leaning model.
	$\textbf{b}$. The workflow of an \textit{anomaly detection} method for force sensing. It involves the utilization of a generative machine learning model comprising a generator and a discriminator for digitally replicating the experimental time-of-flight images.
	 The anomaly detection is assisted by introducing an additional encoder. All three components are deep artificial neural networks. The anomaly score (AS) is composed by residual loss (RL) between input data $x$ and its digital replica $\tilde{x}$ and  discrimination loss (DL) generated by the discriminator. 
}
\label{figure1}
\end{figure}

	\textbf{\textit{The Bose-Einstein condensate system.}}  
We create a bosonic quantum system by trapping about $2\times 10^5$ $^{87}$Rb atoms. This system forms a  BEC at low temperature, about $50$ nK in our experiment (Fig.~\ref{figure1}). The atomic BEC is loaded in a triangular optical lattice to suppress unwanted real-space dynamics~\cite{GUO2022Bulletin}. After preparation of the lattice BEC system, the optical lattice and the trapping potential are shut-off, which then let the atoms expand ballistically. Performing the time-of-flight (TOF) {technique~(see Supplementary Note 2 \cite{SuppM}) in the experiment}, we measure the momentum distribution $n({\bf k})$. It takes about $T_0 = 38$ s to complete one experimental cycle.  {Details of our experimental setup can be found at Ref.~\cite{GUO2022Bulletin}.}

In the time-of-flight measurements, we have shot-to-shot noise. There are quantum shot-noises, which arise from quantum superposition of different momentum eigenstates due to atomic interactions, trapping potentials, and optical lattice confinement. 
There are thermal noises. Although the atomic BEC is cooled down to $50$ nK, we still have thermally activated atoms. These atoms induce stochastic fluctuations in the measurement outcomes. 
There are also technological noises. The control of the atom numbers, the trapping potential, and the optical lattice depth are not perfect in the experiment.  They may have noticeable or unnoticeable drifts in different experimental runs. 
As a result, the TOF measurement outcomes in the consecutive experimental runs would unavoidably fluctuate from shot to shot. We thus denote the measurement outcomes  as $n_\alpha ({\bf k})$, with $\alpha$ indexing different experimental runs.

In detecting an external force acting on the BEC, a standard approach is to examine the response in the averaged center-of-mass (COM) momentum, i.e., 
\be 
\overline{\bf k}_{\rm COM} =\frac{ \sum_{\alpha}  \int d^2 {\bf k} n_\alpha({\bf k}) {\bf k} } {\sum_{\alpha} \int d^2 {\bf k} n_\alpha({\bf k})} . 
\label{eq:kcom} 
\ee  
Although the measurement outcome $n_{\alpha} ({\bf k})$ is a two-dimensional image, which potentially involves very rich information, 
the conventional approach of data processing following Eq.~\eqref{eq:kcom} only consumes the zeroth and first order moments of the two-dimensional data, leaving behind higher-order correlations.

\textbf{\textit{The digital replica.}} 
In order to incorporate the full information in the measurement outcomes, $n_{\alpha} ({\bf k})$, one plausible way is to perform digital twinning of the physical system by matching the probability distributions. It then automatically takes into account high-order correlation effects.  
Since the atomic BEC system in the experiment has noises of various origins, it is impractical to simulate the experimental measurement outcomes using conventional  modeling approaches, for example by simulating Gross-Pitaevskii equations~\cite{1961_Gross,1961_Pitaevskii}. 
In this study, we create a digital twinning of the experimental system, by implementing a generative machine learning model, which incorporates quantum, thermal, and technical noise channels simultaneous at equal-footing in a purely data-driven approach.

\begin{figure}[htp]
		\includegraphics[width=1\linewidth]{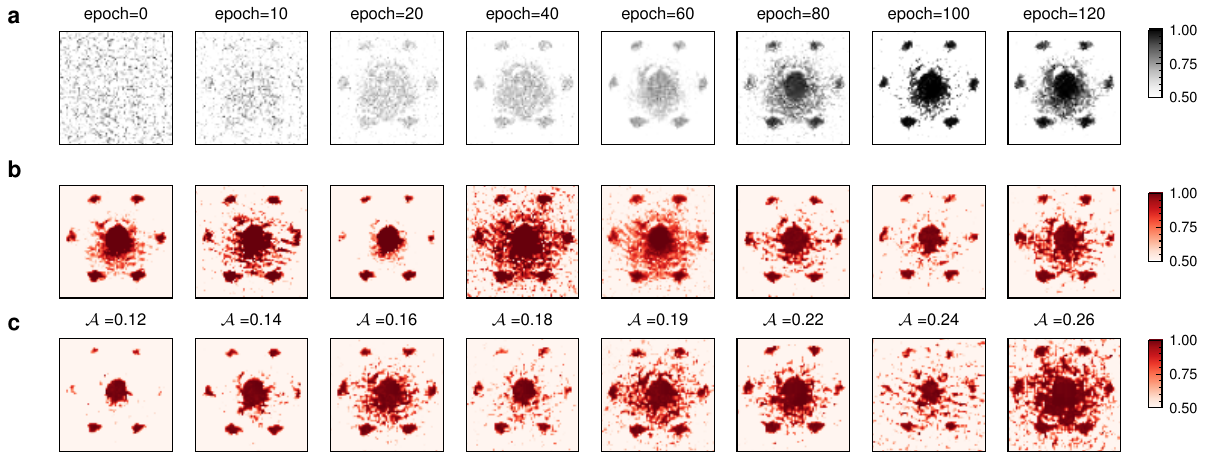}
		\centering
		\caption{ \textbf{The digital replica of time-of-flight  measurements}. {\bf a}. The momentum distribution constructed by generative machine learning model at different training epoch. We present several images generated from a fixed generator, namely the digital replica in {\bf b}. 
{\bf c} shows eight representative real time-of-flight images with anomaly scores from $0.12$ to $0.26$. 
	} 
		\label{figure2}
\end{figure}

We implement a generative adversarial network (GAN)~\cite{Goodfellow2016DL} for digital twinning of the atomic BEC. 
GAN consists of a {generator} $G(\cdot)$ that attempts to map  a latent vector ${\bf z}$ to a realistic data of momentum distribution, 
$G(\cdot)$: ${\bf z}\mapsto \tilde{n} ({\bf k})$,
and a {discriminator} $D(\cdot)$ {that tries}  to differentiate the real data, $n({\bf k})$, from the fake data from the generator, $\tilde{n} ({\bf k}) = G({\bf z})$. 
The two networks are trained simultaneously, with the generator attempting to produce data that can fool the discriminator, and the discriminator learning to correctly identify synthetic data (Methods). 
The generator and  discriminator are realized by two parameterized deep neural networks, denoted as  $G({\bf z} ;\theta_G)$ and $D(n({\bf k}) ;\theta_D)$. 
We collect $3.6 \text{k}$ independent measurements of momentum distributions for force-free experiments, and feed {them} to GAN~(see Supplementary Note 3 \cite{SuppM}).
This amount of data takes about forty hours to collect in experiments, which is {reasonably affordable.}  
In Fig.~\ref{figure2}.{\bf a}, we present fake data produced by the generator during training procedure, where the generated data is visually similar to real data, indicating that the model is capable of capturing the underlying data distribution without model collapsing.
The trained generator produces a digital replica {(Fig.~\ref{figure2}.{\bf b})} of the experimental measurements {(Fig.~\ref{figure2}.{\bf c})}, which could generate fluctuating configurations involving all noise channels automatically.

\begin{figure}[h]
		\centering
		\includegraphics[width=1\linewidth]{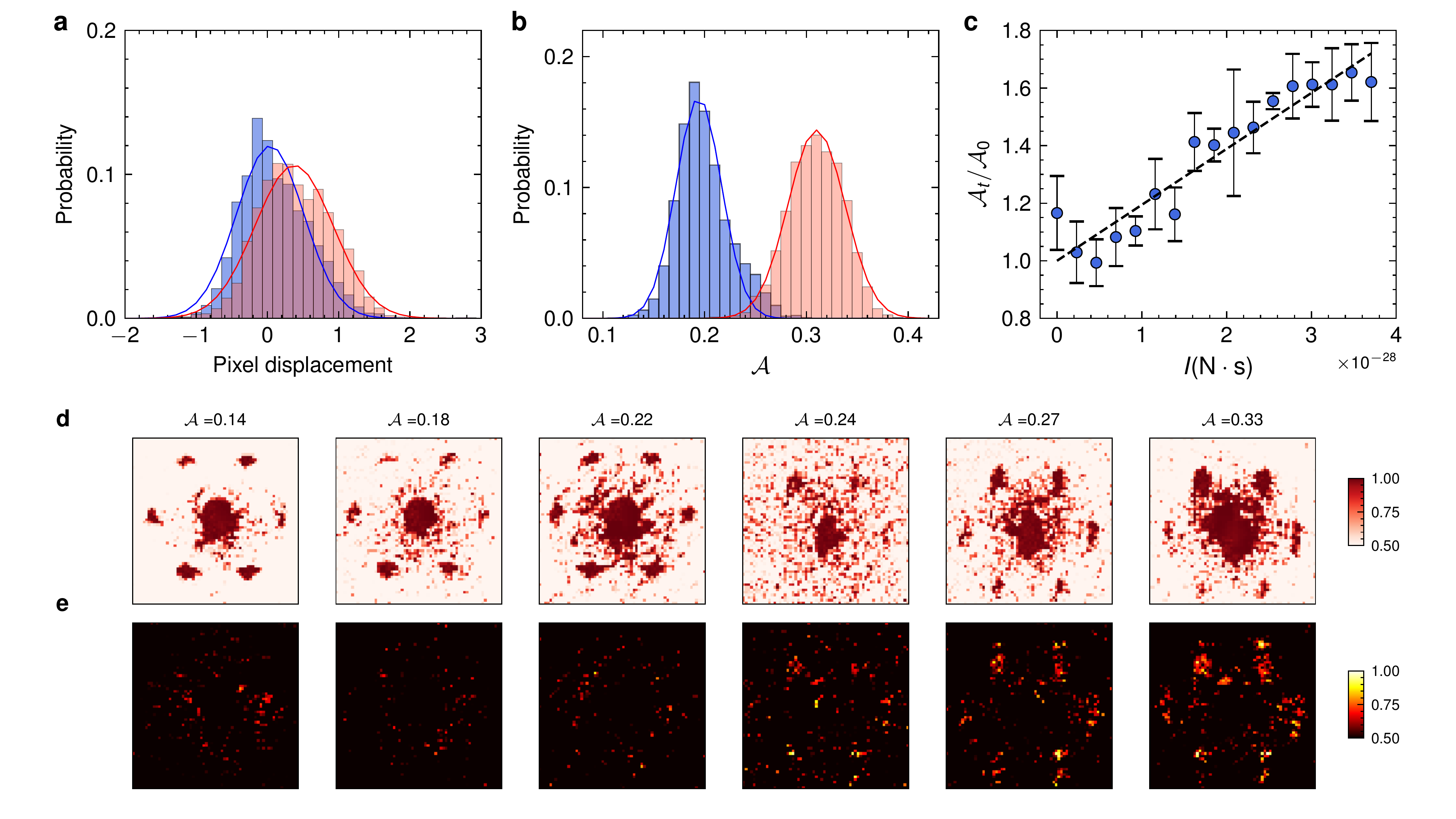}
	\caption{\textbf{Sensing with anomaly score.} \textbf{a-b}. The probability distribution of averaged center-of-mass (COM) momentum $\overline{\bf k}_{\rm COM}$ (in the unit of pixels) and anomaly score $\mathcal{A}$ in the absence of force (blue) and in the presence of force (red), respectively. Noting that the pixel displacement refers to the distance between COM and the y-axis center of pixel-wise images.
		\textbf{c}.The dimensionless anomaly score as a function of impulse $I=F_0\cdot \Delta T$ for an identical optical force $F_0$. 
		The normalization factor in {\bf c} (${\cal A}_0$) is the averaged anomaly score in absence of the external force. The error bar in {\bf c} represents one-sigma statistical uncertainty over six data points.
		The  anomaly score $\mathcal{A}_t(\Delta T)$ corresponds to signal-involved experiments with a force acting  on the Bose-Einstein condensates for a time duration $\Delta T$.
		{\bf d} and {\bf e}, show  the real time-of-flight images and their corresponding anomaly localization ({\it see the main text}), respectively.   Noting that the left (right) three instances in {\bf d} are sampled from  datasets with force-free (force-involved) environments. {More technical details can be found in Method and Supplementary Note 2 and 3.} 
	}
		\label{figure3}
	\end{figure}

As in the experimental data that has typical and atypical configurations due to noises, the digital replica also generates typical and atypical configurations. 
Within the framework of GAN, the degree of atypicality is quantified by an anomaly score,  
\begin{equation}
\label{AS}
	\mathcal{A}( n({\bf k}) )= \mathcal{A}_{R}( n({\bf k}) )+\lambda \cdot \mathcal{A}_D ( n({\bf k}) ). 
\end{equation} 
{The anomaly score combines both residual loss and discrimination loss, serving to quantify the atypicality of an input image $n({\bf k})$ in comparison to its corresponding digital replica $\tilde{n}({\bf k})$.}
Specifically, the residual loss (RL) ${\cal A}_R  =  \| n({\bf k}) - \tilde{n}({\bf k}) ) \|_2 $ is produced by adding an encoder in front of the generator (Fig.~\ref{figure1}),  assessing the Euclidean distance between the real image and its corresponding digital replica. The discrimination loss (DL) $\mathcal{A}_D(n({\bf k})) = \|h({n({\bf k})})-h(\tilde{n}({\bf k}))\|_2 $ is directly given by the feature layer $h(\cdot)$ of the discriminator~(see Supplementary Note 4 \cite{SuppM}).
The encoder is trained by minimizing the residual loss.
The weighting coefficient  $\lambda\in\mathbb{R}$ is a hyper-parameter that balances RL and DL.
 These two components evaluate the discrepancy between fake and real data in terms of image distance  and feature discrimination. 
{We choose $\lambda = -0.76$ for an optimal sensitivity~(see Supplementary Note 3 \cite{SuppM}). For a more in-depth understanding of  Eq.(\ref{AS}), additional detailed interpretations can be found in Method and Supplementary Materials~\cite{SuppM}.}
 We find that the anomalous score has an approximate normal distribution (Fig.~\ref{figure3}.{\bf b}), which indeed reflects the typicality of momentum distribution data.

\textbf{\textit{Sensing by anomaly detection.}} 
When an external force is applied, the physical BEC system then produces TOF data $n({\bf k})$ of different distributions. Conventional sensing schemes examine the response in the COM momentum (Eq.~\eqref{eq:kcom}). The distributions of the COM momentum with and without an external force are different. Force sensing requires differentiating such distributions. When a weak force ($F_0=7.81\times10^{-26}~\text{N}$  in our experiment) is applied, the distribution of COM momentum is only very weakly affected, with a barely noticeable difference from force-free distribution.

With the digital replica, namely the generative machine learning model, we compute the anomaly score ${\cal A}(n({\bf k}))$, a highly nonlinear function of $n({\bf k})$, that could incorporate higher order correlations of the data. This provides a systematic nonlinear data processing approach, much different from the linear data processing as in analyzing the simple COM momentum. 
As a result, the anomaly score is much sensitive to the force {as it effectively reflects high-order information of the experimental data.} 
The resultant distribution of the anomaly score caused by the weak external force is significantly different from the force-free distribution, in sharp contrast to the COM momentum (Fig.~\ref{figure3}.{\bf a}). 
Despite the nonlinearity in the data processing, the response of the anomaly score remains to be linear to the external force (Fig.~\ref{figure3}.{\bf c}).

By comparing the distributions of the COM momentum and the anomaly score, it is evident that analyzing the anomaly score, known as {\it anomaly detection} in the context of machine learning, is more efficient for detecting the external force applied to the BEC system.

{
We further investigate the primary characteristics that contribute to the anomaly score in presence of an external force.
Specifically, we assess the momentum dependence of the residual loss, ${\cal A}_R$, namely, 
\be
n_R ({\bf k})  =  | n(\textbf{k})-G^*(E^*(n(\textbf{k}))) |, 
\ee
with the generator $G^*(\cdot)$ and the encoder $E^*(\cdot)$ being fixed.
In Fig. \ref{figure3}.{\bf d}, we provide 
six representative atomic images from real experimental datasets labeled with anomaly scores. 
 {In Fig.~\ref{figure3}.{\bf e}, we observe bright speckles in $n_R({\bf k})$(commonly referred to as anomaly localization~\cite{schlegl2017AnoGAN}), with these speckles becoming increasingly prominent as the applied force intensifies. 
 This implies the signals contributing to the force sensing mainly come from the large momentum region in the second Brillouin zone. 
{The anomaly localization is  consistent with the common understanding in the BEC community 
that the momentum distribution at small momenta (near zero) is more affected by interaction effects and imaging saturation artifacts, 
indicating the machine learning model successfully captures the relevant features of the experimental data.}}
Further discussion on model interpretability from the perspective of feature representation \cite{schlegl2017AnoGAN,f-AnoGAN,tangyou22ML} is provided in Supplementary material \cite{SuppM}.
Our findings regarding anomaly localization, exemplified by the presence of a hexagonal peak structure in the rightmost portion of Figure \ref{figure3}.{\bf e}, imply that the dominant signal for force sensing originates primarily from high-momentum peaks observed in time-of-flight (TOF) measurements. 
This could be attributed to the reduced impact of various sources of noise, such as atomic scattering, trapping potential, and thermal activation, on the high-momentum components of the Bose-Einstein condensate (BEC), owing to energy separation.
}

		\begin{figure}[htp]
		\centering
		\includegraphics[width=1\linewidth]{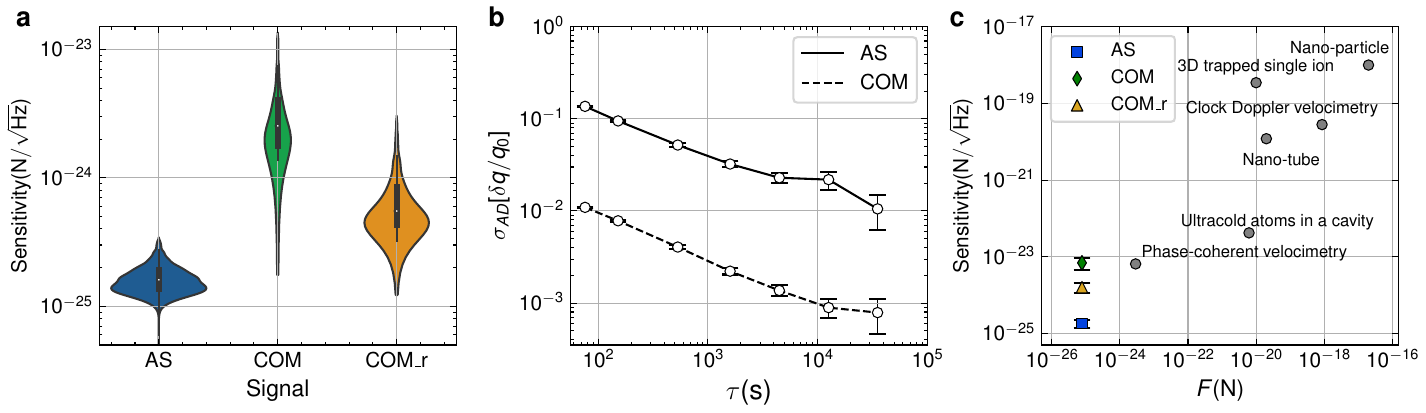}
		\caption{\textbf{Sensitivity and stability of anomaly detection.}
			 \textbf{a}. The sensitivity distribution of using the anomaly score (blue), center-of-mass (COM) momentum with raw data (green) and reduced datasets (yellow), respectively.  {In \textbf{b}, the Allan Deviations $\sigma_{AD}[\delta q/q_0]$ correspond to the anomaly score (AS) and the COM momentum. The momentum difference  $\delta q = q_t-q_0$ represents the signal response,  where $q_0$ is the force-free signal and $q_t$ is the signal  corresponding to the momentum distribution produced by acting the force on Bose-Einstein condensate (BEC).}  Both of them decay with the integration time $\tau$, as $1/\tau$. 
			 In \textbf{c}, we compared related works in sensitivity with force $F$, including measurements based on nano-tube \cite{moser2013ultrasensitive}, nano-particles \cite{Hebestreit2018prl}, trapped ions with phase-coherent velocimetry \cite{biercuk2010ultrasensitive}, a single trapped ion with optical clock Doppler velocimetry \cite{shaniv2017quantum}, a 3D-trapped single ion \cite{blums2018single}, and ultracold atoms in a cavity \cite{schreppler2014optically}. 
			 Note that error bars indicate the one-sigma statistical uncertainty in  panel (\textbf{b}) and (\textbf{c}). }
		\label{figure4}
	\end{figure}

\textbf{\textit{Sensitivity and Stability.}} 
{In order to quantitatively characterize the advantage of the anomaly detection over conventional approaches, we compute the corresponding sensitivity. 
A general force sensing process involves a force $F$ to be detected, and a signal $q$ that is directly or indirectly measured in the experiment. 
{In the previous work~\cite{GUO2022Bulletin}, $q$ corresponds to COM momentum $\overline{\bf k}_{\rm COM}$ (Eq.~\eqref{eq:kcom}). It corresponds to the anomaly score $\mathcal{A}$ (Eq.~\eqref{AS}) in the machine learning approach reported here. }
The measured signals $q$ would fluctuate in consecutive experimental measurements due to noises of various channels. 
We assume 
the fluctuations in different measurements are completely independent.
The strength of the fluctuations is quantified by the standard deviation of $q$, to be referred to as $\sigma_0$. 
A single measurement has a fixed time cost of $T_0$. 
The minimum force we can resolve with one single-measurement is given by 
$V_{\rm min} \sim \sigma_0/|\partial_V q|$. 
Performing $N$ times of experimental measurements, the signal to noise ratio (SNR) is given by $\text{SNR} = \sqrt{N} \times |\partial_V q V|/\sigma_0$.  The one-sigma sensitivity is defined as~(see Supplementary Note 1 \cite{SuppM})
\be 
\label{eq:sensitivity} 
{\cal S} = \sqrt{T_0} \times \frac{\sigma_0}{|\partial_V q|}. 
\ee 
This definition applies to both conventional approaches and the anomaly detection.

Acting linear transformation on $q$, the sensitivity remains the same according to Eq.~\eqref{eq:sensitivity}. However, different signals that are related by nonlinear transformations do not necessarily have the same sensitivity.  
We compare the sensitivities of the COM momentum and the  anomaly detection approaches taking exactly the same set of experimental data. 
For the COM momentum approach, we obtain a sensitivity  $\mathcal{S} ^{\text{COM}} =6.8(9) \times 10^{-24}~\text{N}/\sqrt{\text{Hz}}$ (Fig.~\ref{figure4}.{\bf a}). 
For the anomaly detection, we have $S ^{\text{AS}}=1.7(4)\times10^{-25}~\text{N}/\sqrt{\text{Hz}}$. 
This means the anomaly detection is about $40$ times more sensitive than the COM momentum approach. 

We emphasize that in the above comparison we use the raw experimental data without invoking any prior knowledge of the physical process. The anomaly detection approach is thus entirely data-driven. 
We further examine how much the conventional COM momentum approach can be improved by machine learning based noise reduction. We perform Gaussian processing prior to extracting the COM momentum~(see Supplementary Note 5 \cite{SuppM}). The resultant sensitivity can be improved to $ S_{r}^{\text{COM}}=1.6(4)\times10^{-24}~\text{N}/\sqrt{\text{Hz}}$ (Fig.~\ref{figure4}.{\bf a}). But this is still one-order-of-magnitude worse than our anomaly detection.

In Figure~\ref{figure4}.{\bf c}, we show the comparison of the achieved sensitivity of digital twinning atomic BEC with previous experiments, 
{including phase-coherent velocimetry~\cite{biercuk2010ultrasensitive}, cold atoms in a cavity~\cite{schreppler2014optically}, and trapped ions~\cite{blums2018single}.} 
Our achieved force sensitivity shows orders-of-magnitude improvement over other experiments. 
It is worth to mention that the digital twinning  and the anomaly detection techniques developed here are quite generic. These techniques 
are readily applicable to improve the sensitivity of other experimental setups as well.

For quantum force sensing, it is also important to have long-term stability besides the high sensitivity. This is captured by the Allan Deviation~\cite{overlapAD}, 
which is widely used to examine long-term drifts. 
We confirm that the Allan Deviation of the anomaly detection falls off with the integration time ($\tau$) as $1/\sqrt{\tau}$, having the same scaling as the Allan Deviation of the COM momentum (Fig.\ref{figure4}.{\bf b}). It is thus evident that  no long-term drifts are induced by the nonlinear data processing in anomaly detection. 
The $1/\sqrt{\tau}$ decay also implies the fluctuations of the anomaly score are mainly white noise, which justifies the above definition of sensitivity.

It is worth noting here that the sensitivity can also be enhanced by choosing the high data-quality region of the TOF measurements according to the impulse theorem, as used in a previous study~\cite{GUO2022Bulletin}. Such analysis requires certain prior information of the force, and gives a sensitivity comparable to the present anomaly detection approach. Nonetheless, we emphasize that the anomaly detection approach is purely data driven, and is consequently more robust against long-term drifts in experiments. The improvement in the long-term stability is evident by comparing the Allan Deviation of the anomaly detection to the pervious study~\cite{GUO2022Bulletin}. 
In the previous study, the Allan Deviation bends up at a time scale of $\tau= 10^4$~s, whereas it keep decreasing following $1/\sqrt{\tau}$ even at the time scale of $4\times 10^4$~s.

	\medskip 
	\section*{Discussion}\label{sec12}

In this study, we present an unique method for quantum force sensing using digital twinning of atomic BEC and anomaly detection facilitated by a generative machine learning model. By incorporating complex correlation effects present in the experimental data through nonlinear processing, we achieve a significant enhancement in sensitivity while maintaining long-term stability. Unlike conventional approaches that rely on extracting basic statistical moments from high-dimensional data, such as time-of-flight (TOF) measurements in BEC, our anomaly detection approach employs a neural network-based nonlinear function, denoted as $f_{\rm NN} ({\bf x})$, to fully exploit the information within the high-dimensional data. Through extensive training and iterative refinement, this methodology effectively amplifies the signal-to-noise ratio, as confirmed by convergence in sensitivity with increasing training data~(see Supplementary Note 6 \cite{SuppM}). 

Our findings reveal an intriguing aspect: the sensitivity of a physical sensor is intimately tied to the data processing strategy, denoted as ${\cal S} [f_{\rm NN}]$. This implies the existence of an upper bound, 
\be 
{\cal S}_{\rm opt} = \max_{f_{\rm NN} } \left\{    {\cal S} [f_{\rm NN}] \right\}, 
\ee 
which represents the maximum achievable sensitivity attainable by a given sensor configuration. 
{It is worth mentioning that the role of the digital twin in our framework mimics data behavior in sensing responses, which could be further advanced by a viable generative machine-learning model.}
The determination of this upper bound warrants further investigation in future research endeavors. 

The other important direction is to investigate the fundamental quantum limits of the anomaly detection approach. How the optimal sensitivity ${\cal S}_{\rm opt}$ is fundamentally limited by the quantum shot-noise, and  its scaling with  the number of atoms in the BEC or  the imaging resolution, are worth further theoretical studies.

\medskip 
\section*{Methods}\label{sec11}

	\medskip 
	\textbf{ Generative Adversarial Networks}

 Our primary objective is to leverage Generative Adversarial Networks (GANs) as generative models to construct digital replicas of our experimental data ${\bf x}\sim \mathcal{P}_0({\bf x})$. To begin, we collect a set of raw observables $X= \{{\bf x}_1,{\bf x}_2,{\bf x}_3,...,{\bf x}_N\}$ obtained from $N$ independent experiments.
 In our force sensing setup, we have 3,600 two-dimensional time-of-flight (TOF) atomic images from independent experimental outcomes. 
 {90$\%$ of the signal-free images were utilized for training the generator through adversarial learning, while the remaining data was reserved for testing our model.} Each pixel-wise image is converted to a resolution of $64\times 64$ with pixel normalization within the range of $[-1,1]$. The discriminator $D(\theta_D)$ is constructed using six layers of fully-connected neural networks, with Leaky ReLU activation applied to each layer. The discriminator is trained to produce a one-dimensional scalar $p\in[0,1]$, which represents the probability that the input image ${\bf x}$ corresponds to real data.
 The generator $G(\theta_G)$ consists of five blocks of neural networks, with each block comprising a fully-connected layer, a batch normalization layer, and a ReLU unit. The input dimension of the generator is determined by the size of the latent vector ${\bf z}$, which is set to one hundred in our setup. 
In an adversarial training strategy, the parameters ${\theta_D, \theta_G}$ are simultaneously optimized using a standard min-max loss function, denoted as $V(D,G)$ \cite{Goodfellow2014GAN}:
	\begin{equation}
		\min_G\max_D V(D,G) = \mathbb{E}_{{\bf x}\sim \mathcal{P}_{0}}[\log D({\bf x})]+\mathbb{E}_{{\bf z}\sim \mathcal{P}_{\bf z}}[\log (1-D(G({\bf z})))].
	\end{equation}
The convergence criterion is that the generator is able to successfully deceive the discriminator, leading to the generation of high-quality digital replicas of the training data.

\medskip 
 	\textbf{ Anomaly score } 
 	
The anomaly score consists of two components.
 Firstly, we compute the \textit{residual loss} (RL), denoted as $\mathcal{A}_{R}({\bf x}) = \|{\bf x}-\tilde{{\bf x}}\|_2$, which represents the Euclidean distance between the real image and the fake image generated by the GAN. This loss directly captures the disparity between real and fake images.
 Secondly, we evaluate the \textit{discrimination loss} (DL) of a feature layer, denoted as $\mathcal{A}_D({\bf x}) = \|h({\bf x})-h(\tilde{{\bf x}})\|_2$, where $h(\cdot)$ represents the feature layer of the discriminator. The discrimination loss quantifies the discrepancy between the features extracted from real and fake data \cite{SuppM}. Specifically, we focus on the last fully-connected layer of the discriminator as the designated feature layer.

It is important to note that GANs learn a mapping from random noise ${\bf z}$ to an image drawn from a healthy distribution. However, due to the non-injectiveness of this mapping, multiple random inputs ${\bf z}$ can generate the same fake image. Statistically, this characteristic introduces uncertainty when scoring image anomalies, thus resulting in diminished sensitivity when employing the anomaly score as a signal in the sensing procedure.
To address this issue and establish a more stable signal characterized by a bijective mapping, we introduce an additional neural network called \textit{encoder}. This encoder, denoted as $E(\cdot)$, is trained to inversely generate the latent vector ${\bf z}$ given an input image ${\bf x}$ based on pre-trained GANs. The encoder architecture is designed to be the inverse of the generator. All three fixed networks, namely $G^*(\cdot)$, $D^*(\cdot)$, and $E^*(\cdot)$, are solely trained using the signal-free dataset $X$. Consequently, the anomaly score can be evaluated by leveraging the three fixed deep networks in the following formulation: $\mathcal{A}( {\bf x} )= \|{\bf x}-G^*(E^*({\bf x}))\|_2+\lambda \|h({\bf x})-h(G^*(E^*({\bf x})))\|_2$, where $\lambda$ denotes a weight coefficient.
To determine the optimal weight coefficient that maximizes sensitivity $\mathcal{S}$ as defined in Eq.(\ref{eq:sensitivity}), an extensive search for $\lambda$ is performed within the range $[-1,1]$, aiming to identify the value that yields the highest sensitivity $\lambda_{\text{op}}=\text{argmin}_{\lambda\in[-1,1]}\mathcal{S}(\lambda)$. In our force sensing application, the optimal value was found to be $\lambda_{\text{op}}=-0.76$~(see Supplementary Note 3 \cite{SuppM}).
For further details on model training and evaluation, we refer the reader to the supplementary materials \cite{SuppM}.

\medskip 
\textbf{\bf Sensitivity of Anomaly Detection}

Once the model of GAN is trained, it is used to produce an anomaly score ${\cal A} = f_{\rm NN} ({\bf x})$. The experimental measurement outcome ${\bf x}$ satisfies a certain probability distribution $P_F ({\bf x})$, depending on the external force $F$, {i.e., the physical quantity $V=F$ in Eq.\eqref{eq:sensitivity}}. Although the form of the distribution is unknown, we find the anomaly score calculated according to the experimental data obeys a Gaussian distribution (see Fig. \ref{figure3}.{\bf b}) with a mean value $\overline{ {\cal A}} (F)$, and  a standard deviation $\sigma_0$. We confirm that $\overline{\cal A}$ is linear to the force strength. It is more convenient to adjust the impulse on the BEC by controlling the force acting time than controlling the force directly, so we adjust the force acting time to mimic a variable force according to the momentum-impulse theorem. In Fig.~\ref{figure3}.{\bf c}, we demonstrate the linear scaling of the anomaly score to the impulse. The SNR with one cycle of TOF measurement is 
${\rm SNR} = \overline {\cal A}(F)  /\sigma_0.$  
For the linearity of the response of the anomaly score to the force, the sensitivity of the anomaly detection is 
${\cal S} = \sqrt{T_0} \times 
\left. \frac{\sigma_0}{\partial_F \overline {\cal A}(F) } \right |_{F \to 0} $. 
In terms of the probability distribution, the sensitivity is given by 
\be 
{\cal S} = \sqrt{T_0} 
\left. 
\frac{\sqrt{\sum_{\bf x}   P_F( {\bf x} ) f_{\rm NN}^2 ({\bf x}) - \left[ \sum_{\bf x}    P_F( {\bf x} ) f_{\rm NN} ({\bf x}) \right]^2   } } {\sum_{\bf x} \partial_F P_F ({\bf x}) f_{\rm NN} ({\bf x}) } \right|_{F\to 0}. 
\ee 
It is then apparent that the optimal choice of anomaly score to maximize sensitivity depends on the probability distribution. 
The digital twinning technique provides  a purely data-driven approach to model the distribution, 
which permits the single-parameter ($\lambda$) optimization of the sensitivity. 
We remark here that there is still a large window  for improving the sensitivity, because of the large degrees of freedom in constructing the anomaly score. 

\medskip 	
\vspace{10pt}
\begin{normalsize}
	\noindent
	\textbf{Acknowledgments}
\end{normalsize}
We acknowledge helpful discussion with W. Vincent Liu. 
This work is supported by National Program on Key Basic Research Project of China (Grant
No. 2021YFA1400900), National Natural Science Foundation of China (Grants No. 11934002, 12075128, T2225008), Shanghai Municipal Science and Technology Major Project (Grant No. 2019SHZDZX01), and Shanghai Science Foundation
(Grants No.21QA1400500).

\vspace{10pt}
\begin{normalsize}
    \noindent
    \textbf{DATA AVAILABILITY}
\end{normalsize}

\noindent
Data are available from the authors upon reasonable request.

\vspace{10pt}
\begin{normalsize}
    \noindent
    \textbf{CODE AVAILABILITY}
\end{normalsize}

\noindent
The computation code for producing the results in this work is available upon reasonable request.

\vspace{10pt}
\begin{normalsize}
    \noindent
    \textbf{AUTHOR CONTRIBUTION}
\end{normalsize}

\noindent
X.L. conceived the main idea in discussion with
T.H., Z.N. and X.Z..
 T.H. designed the machine learning framework and carried out the tests. 
Z.Y. contributed to the experimental data analysis. 
 All authors contributed to writing the paper. 

\vspace{10pt}
\begin{normalsize}
    \noindent
    \textbf{COMPETING INTERESTS}
\end{normalsize}

\noindent
The authors declare no competing interests.


\end{document}